\documentclass[reprint,amsmath,amssymb,aps]{revtex4-2}
\usepackage{graphicx}
\usepackage{dcolumn}
\usepackage{bm}

\usepackage{caption}

\usepackage{lipsum}
\usepackage{siunitx}
\usepackage{graphicx}
\usepackage{subcaption}

\begin{document}
\preprint{APS/123-QED}

\title{Analysis of Skew Quadrupole Compensation in RF-Photoinjectors}

\author{P. Denham\email{pdenham@physics.ucla.edu}, F. Cropp, P. Musumeci}%
\affiliation{%
 University of California at Los Angeles, Los Angeles, California, USA
}%
\thanks{Work supported by the US National Science Foundation under grant DMR-1548924, PHY-1549132 and PHY-1734215}

\begin{abstract}
In this paper, we present a detailed analysis of the effects of  quadrupole components in radiofrequency (RF) photoinjectors, which could occur either in the RF cavity or in the gun solenoid, and their compensation using a skew quadrupole corrector. The analysis is generalized to the case in which the correction cannot be applied at the same location as the quadrupole. A practical guide to setting the corrector to eliminate correlation in the x-y space is discussed. Particle tracking simulations and measurements of emittance and beam matrix elements obtained at the UCLA Pegasus photoinjector are presented to validate the analysis.
\end{abstract}

\maketitle

\section{Introduction}
The phase space density of modern electron sources, a fundamental figure of merit for beam quality, continues to increase due to the advances in our understanding of the photoemission physics at the cathode as well as improvement in the gradient of electron guns \cite{Musumeci_advances, Cultrera_AA, akre_lclsinjector, rosenzweig_topgun, PSIcommissioning}. Consequently, it becomes important to make sure that the transport line after the injector preserves the initial beam quality from the photocathode and limits the emittance dilution effects caused by space charge forces or other non-linearities \cite{dowell_emitsources, Zhou_transversenonuniformities}. Correlations between the different beam phase spaces (either transverse-longitudinal, or transverse-transverse) can also cause significant beam degradation, but they can often be reversed by proper beam dynamics transformations (for example emittance compensation   \cite{carlsten_emitcomp}) to unwind the correlations and recover the original emittances. In this paper, we focus our attention on one such source of projected emittance growth, the x-y coupling induced by rotated quadrupole components. Quadrupole fields are linear and do not cause any increase in the total 4D phase space volume, but have the potential to significantly dilute the emittance projections along $x$ and $y$ through the introduction of unwanted correlations in the beam phase space distribution.

Rotated quadrupole fields typically arise either due to the asymmetries of the gun RF feeding port, if a properly symmetrized coupler design is not employed \cite{gunquadrupole} or due to small imperfections in the coils or yoke of the emittance compensation solenoid. Even in the case where the unwanted (error) quadrupolar fields are x-y aligned before the solenoid, the Larmor rotation will skew the beam distribution after the gun.

The effects of skew quadrupoles acting on the beam distribution are well documented \cite{Dowell_sqc}, and widely known to impact significantly the transport and diagnosis of high brightness beams \cite{Zheng_thermalemittancecoupled, Krasilnikov:SQC}. Experimental investigation in RF photoinjectors as well as the possibility of correcting the induced correlations has been recently discussed in the literature \cite{Zheng_correction}. In the treatment offered in these papers, one would want to place the corrector element right at the plane of the error. Nevertheless this is often impossible, either because the error plane is not fully known or because it is not easily accessible in the dense beamline arrangements typical of photoinjectors.

Here we follow up and complement the existing analysis extending it to the case when the error quadrupole and the corrector are not in the same plane and an arbitrary phase advance exists between them. A general formula for the strength of the corrector is obtained and the conditions to obtain a perfect compensation are derived. Particle tracking simulations as well as phase space measurements taken at the UCLA Pegasus advanced photoinjector laboratory are used to support our studies. Finally, a general guidance principle on how to set a corrector without the need to measure in detail the beam phase space distribution is offered and justified.

The paper is organized as follows: we first review the zero phase advance case to set up the problem and identify the physical quantities that play a role in the projected emittance growth. We then discuss the compensation process and generalize it to the case of arbitrary phase advance. It is found that for a $\pi$ phase advance, the possibility of full compensation exists. General Particle Tracer (GPT) \cite{GPT} tracking simulations are used throughout the paper to extend the results of the analytical calculations to the realistic case of the UCLA Pegasus beamline which is taken as a fairly general example of S-band high gradient RF photoinjector. We experimentally validate our analysis using a simple skew quadrupole corrector design recently installed at UCLA. The TEM grid shadowgraphy technique is used to retrieve the second order moments of the full 4D beam matrix and the beam emittances as a function of the corrector settings.

\section{Quadrupole Compensation}

In this section, we analyze the ability to correct the effects of unwanted quadrupole moments that arise in the vicinity of the gun. We start with the case of zero phase advance between the corrector and error quadrupole. Many of the results that we present here have already been discussed in \cite{Dowell_sqc,Zheng_thermalemittancecoupled} and the follow-up paper \cite{Zheng_correction}. Here we generalize to the case of non-zero phase advance and show that an effective compensation can be obtained in general conditions and one can fully compensate the projected emittance growth in the special case that the phase advance between the corrector and the quadrupole error is exactly and integer multiple of $\pi$, i.e., when the particle distribution at the error quadrupole is imaged at the corrector quadrupole.

\subsection{Zero Phase Advance}
We start by calculating the effect of a single rotated quadrupole on each of the projected emittances. In the thin lens approximation, the transport matrix through an error quadrupole of focal length $f_e$ rotated by an arbitrary angle $\alpha$ is given by:
\begin{equation}
R_{f_e}(\alpha)=
\left(
\begin{array}{cccc}
 1 & 0 & 0 & 0 \\
 \frac{\cos (2 \alpha )}{f_e} & 1 & -\frac{\sin(2 \alpha)}{f_e} & 0 \\
 0 & 0 & 1 & 0 \\
 -\frac{\sin(2 \alpha)}{f_e} & 0 & -\frac{\cos (2 \alpha )}{f_e} & 1 \\
\end{array}
\right)
\end{equation}
We are interested in the effects of such a lens on the second moments of the beam distribution function. If the trace space of the beam is initially uncorrelated, then the beam matrix has a block diagonal form $\Sigma_0=\left(\begin{array}{cc}
\Sigma_{XX} & 0_{2x2} \\
0_{2x2} & \Sigma_{YY} \\
\end{array}
\right)
$.
The beam matrix is mapped through the quadrupole by the matrix transport rule, $\Sigma_f=R_{f}(\alpha)\Sigma_0R^T_{f}(\alpha)$. The resulting horizontal geometric (i.e. unnormalized) projected emittance $\epsilon_{x}$ is given by the square root of the determinant of the x-x' trace space block of the final beam matrix. Explicitly, the result simplifies to:
\begin{equation}
\epsilon_{xf}^2=\epsilon_{x0}^2+\frac{\langle xx \rangle \langle yy \rangle \sin ^2(2 \alpha )}{f^2}
\label{emit_equation}
\end{equation}
where $\epsilon_{x0}$ is the horizontal trace space emittance before the quadrupole, $\langle xx \rangle$ and $\langle yy \rangle$ are the second order moment of the beam distribution in the x and y direction respectively. The same expression arises in the y-y' trace space. 

\begin{figure}[]
\includegraphics[scale=0.5]{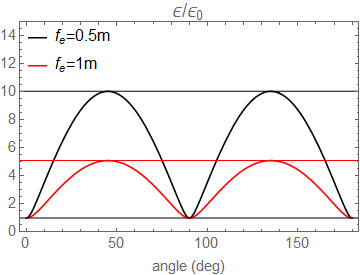}
\captionsetup{justification=centerlast}
\caption{Emittance growth as a function of quadrupole rotation angle for two different focal lengths. The initial geometric emittance is 50 nm and the rms x and y beam size are 0.5~mm. Notice how the projected emittances can increase substantially, severely impacting the downstream beam quality.}
\label{emittance_growth}
\end{figure}

It is clear that when the angle $\alpha\to0$, then the emittance projections are conserved. Furthermore, the effect of the quadrupole field is minimal if the beam is very small as the emittance growth depends on the rms spot sizes at the quadrupole plane. We will see later that this latter feature negatively affects the ability to correct emittance growth when an arbitrary phase advance (and resulting change in spot size) occurs between the error quadrupole and the corrector. In general, the projected trace space area can be drastically altered from the initially uncoupled value when $\alpha\to\pi/4$. A useful quantity is the ratio $\frac{\langle xx \rangle \langle yy \rangle }{f^2\epsilon_{x0}^2}$ because it provides an upper bound on the relative emittance growth scaling. In Fig. \ref{emittance_growth}, the square root of Eq. \ref{emit_equation} is plotted with relevant beam parameters for 0.5~m and 1~m focal length error quads, in the case of a round beam with an rms 0.5~mm spot size, and initial geometric emittance 50 nm (consistent with beam parameters at the exit of the Pegasus gun). Note the horizontal lines represent the upper bound on the emittance growth.
Indeed, placing another quadrupole with rotation angle $\theta$ and focal length $f_c$, immediately after the quadrupole error will change the emittance as:
\begin{equation}
\epsilon_{xf}^2=\epsilon_{x0}^2+
\frac{\langle xx \rangle \langle yy \rangle \left(f_c \sin\right(2 \alpha\left)+f_e \sin\right(2 \theta\left)\right)^2}{f_e^2f_c^2}
\end{equation}
When $ f_c \sin(2 \alpha)+f_e \sin(2 \theta)=0$, then the emittance growth vanishes. The trivial instance of this is when $f_c=-f_e$, and $\theta = \alpha, \pi-\alpha/2$. However, there is an entire range of focal lengths for which the corrector will have two correction angles. Specifically, if the focal length of the corrector satisfies $f_c \leq \frac{f_e}{\sin \left(2 \alpha \right)}$, there will be two correction angles, which both converge to $\pi/4$ at equality. This result is shown graphically in Fig. 2, in which the error quad was chosen to have a 1~m focal length and to be rotated by 30 degrees. Again, a round beam with 0.5~mm rms spot size, 1 mrad rms divergence, and $50$ nm rms emittance was used. As the correction quadrupole angle is adjusted, the beam profile rotates as shown by the variation of the rms beam sizes in the central plot of Fig. 2.

\begin{figure*}[t]
    \centering
    \begin{subfigure}[b]{0.3\textwidth}
        \includegraphics[width=\textwidth]{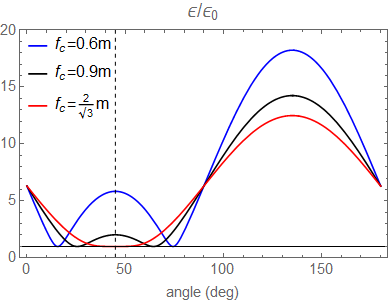}
        \caption{}
        \label{fig:emit}
    \end{subfigure}
    ~ 
    \begin{subfigure}[b]{0.3\textwidth}
        \includegraphics[width=\textwidth]{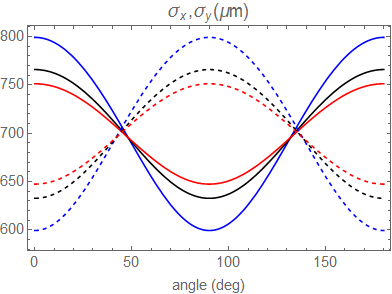}
        \caption{}
        \label{fig:spots}
    \end{subfigure}
    ~ 
    \begin{subfigure}[b]{0.3\textwidth}
        \includegraphics[width=\textwidth]{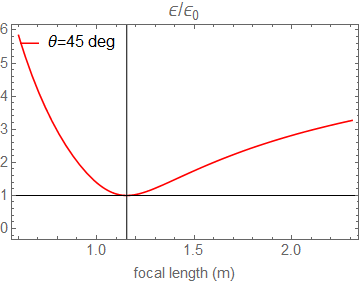}
        \caption{}
        \label{fig:fixedang}
    \end{subfigure}
    \captionsetup{justification=centerlast}
    \caption{(a) Plots of trace space emittance normalized to the initial value before the error quad. The vertical line is positioned at 45 degrees. (b) Rms horizontal (x, solid lines) and vertical (y, dotted lines) spot sizes plotted with quadrupole rotation angle. (c) Normalized emittance plotted against the corrector focal length with the corrector oriented at 45 degrees.}\label{fig:animals}
\end{figure*}


Three different cases corresponding to different corrector focal lengths are shown. The final focal length is tuned to $\frac{f_e}{\sin \left(2 \alpha \right)}$. The vertical line is positioned at 45 degrees. The plots suggest two alternative strategies to find the optimized corrector settings for retrieving the initial projected emittance values. These two strategies correspond either to setting the corrector angle to satisfy $\sin(2 \theta) = -\frac{f_c}{f_e}\sin(2 \alpha)$ for a fixed quadrupole strength or to setting the angle at 45 degrees and dialing the focal length of the corrector to $f_c=\frac{f_e}{\sin \left(2 \alpha \right)}$. 

In either case, the corrected transport through the two quadrupoles simplifies:
\begin{equation}
R=R_{f_c}(\pi/4)R_{f_e}(\alpha)  =  \left(
\begin{array}{cccc}
 1 & 0 & 0 & 0 \\
 -\frac{\cos (2 \alpha )}{f_e} & 1 & 0 & 0 \\
 0 & 0 & 1 & 0 \\
 0 & 0 & \frac{\cos (2 \alpha )}{f_e} & 1 \\
\end{array}
\right)
\end{equation}
Thus, an initially block diagonal second moment beam matrix will remain block diagonal (i.e. uncoupled) through the corrected transport. Unfortunately, this transport breaks round beam symmetry.

\subsection{Finite Phase Advance}
\begin{figure}[]
\centering
  \includegraphics[scale=0.5]{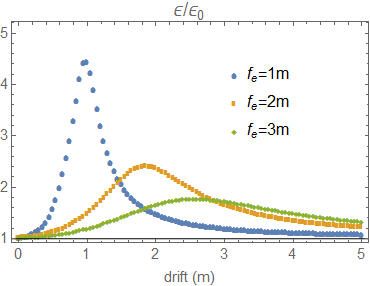}
  \captionsetup{justification=centerlast}
  \caption{The ratios of final corrected projected emittance to input emittance are shown as a function of drift length distance between the corrector and error quadrupole.}
\label{emit_drift}
\end{figure}

When a drift space or an intermediate lens is introduced between the error quadrupole and the corrector, there is a finite betatron phase advance between the two elements and the emittance compensation is no longer exact. This happens because the beam distribution is no longer matched between the corrector and error quadrupole planes. Consequently, the emittance growth can no longer be made identically zero at any correction angle. This effect is shown in Fig. \ref{emit_drift} where the minimum emittance (after the corrector strength and angle are optimized) is plotted as a function of the drift length distance between the error and corrector quadrupoles for three  error quadrupole focal lengths (1~m, 2~m, and 3~m). Note the corrected emittance has a maximum when the drift distance between the quadrupoles is near the focal length of the error quad where the beam has a waist at the corrector plane. As the drift length between the quadrupoles increases, the compensated emittance is nearly restored to the initial value.
The results of Figure 3 suggest that a complete compensation can be recovered when the phase advance between the error quad and the corrector equals $\pi$ as can be directly shown using Courant-Snyder theory.

The solution for motion through an uncoupled quadrupole focusing channel between the quadrupole error and corrector
can be expressed as a symplectic 4x4 uncoupled transport map of the form $M=\left(\begin{array}{cc}
M_x & 0_{2x2} \\
0_{2x2} & M_y \\
\end{array}
\right)
$ where the x-x' trace space mapping is given by:

\begin{widetext}

\begin{equation}
\begin{aligned}
    M_x=\begin{pmatrix}
\sqrt{\frac{\beta_x}{\beta_{x0}}}[\cos(\Delta \psi_x)+\alpha_{x0}\sin(\Delta \psi_x)]& \sqrt{\beta_x \beta_{x0}}\sin(\Delta \psi_x) \\
 -\frac{\alpha_{x0}-\alpha_x}{\sqrt{\beta_x\beta_{x0}}}\cos(\Delta \psi_x)-\frac{1+\alpha_x\alpha_{x0}}{\sqrt{\beta_x\beta_{xo}}}\sin(\Delta \psi_x)& \sqrt{\frac{\beta_{x0}}{\beta_{x}}}[\cos(\Delta \psi_x)+\alpha_{x}\sin(\Delta \psi_x)] \\
\end{pmatrix}
\end{aligned}
\end{equation}
\end{widetext}
in terms of the Courant Snyder parameters $\beta_x$,$\alpha_x$, and the betatron phase advance $\Delta \psi_x$ . $M_y$ has the same form with the subscripts interchanged. When the transport map is known, then tracking the coordinates $u^T=(x,x',y,y')$ is simply a matter of matrix algebra, i.e., $u_f=Mu_i$. An intermediary transport between the corrector and error quadrupoles which simultaneously has a $\pi$ phase advance in each trace space and satisfies $\frac{\beta_{xf}}{\beta_{x0}}=\frac{\beta_{yf}}{\beta_{y0}}=m$ (where $\beta_{xf}, \beta_{x0}, \beta_{yf},$and $\beta_{y0}$ are the initial and final betatron functions for each trace space), results in a transport matrix in an imaging condition of the form:
\begin{equation}
M_{img}=
\begin{pmatrix}
m & 0 & 0 & 0\\
\xi_1 & 1/m & 0 & 0\\
 0& 0 & m & 0\\
 0& 0 & \xi_2 & 1/m
\end{pmatrix}
\end{equation}

In this case, the complete beam matrix transport, including both the error and correction quadrupole yields a projected emittance which can be written as:
\begin{equation}
    \epsilon_{xf}^2=\epsilon_{x0}^2+\frac{\langle xx \rangle\langle yy \rangle \left(\sin (2
   \alpha ) f_{\text{c}}+m^2 f_{\text{e}} \sin (2 \theta
   )\right){}^2}{f_{\text{c}}^2 f_{\text{e}}^2}
\end{equation}
The magnification enters now because the initial particle distribution has been imaged and possibly magnified at the corrector entrance. Thus, the zero phase advance solution is essentially recovered. 

\begin{figure}[]
\centering
  \includegraphics[scale=0.5]{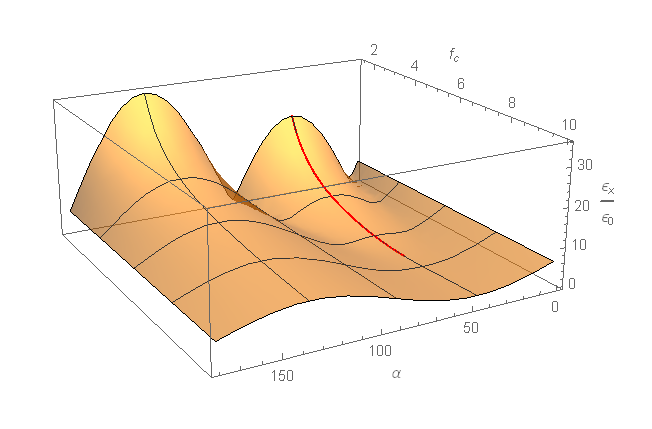}
  \captionsetup{justification=centerlast}
  \caption{Contour plot of the emittance correction parameters space with an intermediary drift space of 1m. The red line corresponds with a parametric
  gradient tuning curve used to minimize the emittance.}
  \label{emit_contour}
\end{figure}

It can be challenging to configure a setup in which the intermediate transport is imaging. In the case where this is not feasible, the emittance correction parameter space still trends as it did when the phase advance was zero. The difference now is that when the emittance is plotted as a function of corrector focal length and angle, the surface is bounded below by the initial, uncoupled emittance with some separation which cannot be made zero. In Fig. \ref{emit_contour} the entire emittance compensation parameter space is shown when there exists a 1~m drift between the error and corrector quadrupoles. It can be seen here that the two correction angles reside on a contour of minimum emittance, just as they do in the zero phase advance case. The correction angles converge to $\pi/4$ as the strength of the corrector is optimized. When the transport is not ideal, one must resort to minimizing the emittance growth by proceeding to optimize the corrector in the same way as the zero phase advance case, i.e., either varying the angle while keeping the strength fixed or the opposite. Intermediary quadrupoles and solenoids do not alter the form of the parameter space, but they do scale it in often inconvenient ways because of accumulated phase advance, which can be most problematic when there is a waist at the corrector.

Incomplete compensation implies the existence of residual coupling between the two trace spaces. It is worth noting that in this case the coupling elements of the beam matrix typically evolve during propagation. For an uncoupled transport (like the drift space case analyzed above, or any up-right quadrupole-based beamline) the final projected emittances are conserved and still given by the expression in Eq. \ref{emit_equation} obtained right after the application of the error quadrupole lens. This can be seen by applying an uncoupled (block-diagonal) transport matrix to the resulting coupled beam matrix.
\begin{equation}
\begin{split}
\Sigma_f&=\left(\begin{array}{cc}
R_{XX} & 0_{2x2} \\
0_{2x2} & R_{YY} \\
\end{array}
\right)
\left(\begin{array}{cc}
\Sigma_{XX} & \Sigma_{XY} \\
\Sigma_{XY}^T & \Sigma_{YY} \\
\end{array}
\right)_i
\left(\begin{array}{cc}
R_{XX}^T & 0_{2x2} \\
0_{2x2} & R_{YY}^T \\
\end{array}
\right) \\
&=\left(\begin{array}{cc}
R_{XX}\Sigma_{XX}R^T_{XX} & R_{XX}\Sigma_{XY}R^T_{YY} \\
R_{YY}\Sigma^T_{XY}R_{XX}^T & R_{YY}\Sigma_{YY}R^T_{YY} \\
\end{array}
\right)
\end{split}
\end{equation}
It immediately follows that the projected emittance in each trace space is conserved in such transformation since, $\epsilon_{x}^2=\det R_{XX}\Sigma_{XX}R^T_{XX}=\det \Sigma_{XX}$.

\label{emit_corrsolenoid}



\section{Experimental tests}
At the UCLA Pegasus beamline we tested some of these concepts in practice. Over the years, small coupling between the horizontal and vertical phase space have been observed on beam profile monitors and in emittance measurements, yet they had not been fully characterized. More recently, as research efforts were focused on the investigation of flat-beam transformation in the ultralow charge regime and generation of very small beams for DLA applications \cite{cesar_DLA}, the control of very small amounts of x-y coupling became crucial and provided the motivation for many of the studies discussed in this paper. 

The RF photoinjector in the beamline is a hard-copper clamped version of the UCLA/SLAC/BNL 1.6 cell S-band photoinjector with dipole-symmetrized RF feeding, but no race-track geometry to cancel the RF quadrupole field components \cite{Alesini_gun}. In addition, magnetic meaurements of the emittance compensation solenoid installed at the gun also revealed small quadrupole components.

\subsection{Skew Quadrupole Corrector Design}
In order to correct the unwanted coupling, we developed a compact Skew Quadrupole Corrector (SQC) that could fit inside the emittance compensation solenoid. The design is based on the field produced from four parallel wires oriented in a quadrupole configuration. The expression for such a configuration is given by the following:
 \begin{equation}
\boldsymbol{B}= \frac{2\mu_0 I}{\pi R^2}\left(y\boldsymbol{\hat{x}}-x\boldsymbol{\hat{y}}\right)
 \end{equation}
Here, $I$ is the current through each of the wires, and $R$ is the radius of the circle through which the wires intersect. Our device was made by taking two of these quadrupole configurations rotated 45 degrees relative to each other and superimposing the fields yields then:
 \begin{equation}
     \boldsymbol{B}=\boldsymbol{B}_{norm}+\boldsymbol{B}_{skew}= \frac{2\mu_0}{\pi R^2}\left[(I_1 y- I_2 x)\boldsymbol{\hat{x}}+(I_1 x+ I_2 y)\boldsymbol{\hat{y}}\right]
     \label{B_currents}
 \end{equation}
To determine the currents $I_1$ and $I_2$ that generate a quadrupole rotated at an angle $\alpha$ with a gradient $G$, we refer to the scalar potential for a quadrupole field oriented normally in the rotated frame with coordinates related to the lab frame by $u=\cos (\alpha ) x + \sin  (\alpha )y$, and $v=-\sin (\alpha ) x + \cos  (\alpha )y$. The potential in the rotated frame is then:
 \begin{equation}
     V=Guv
 \end{equation}
In the vacuum region between the wires, the magnetic field is found from $\boldsymbol{B}=-\nabla V$. In terms of lab coordinates:
 \begin{equation}
     \boldsymbol{B}=G(y\cos (2\alpha )-x\sin(2\alpha))\boldsymbol{\hat{x}}+G(x\cos (2\alpha )+y\sin(2\alpha))\boldsymbol{\hat{y}}
     \label{B_gradient}
 \end{equation}
 
\begin{figure}[]
\centering
  \includegraphics[scale=0.059]{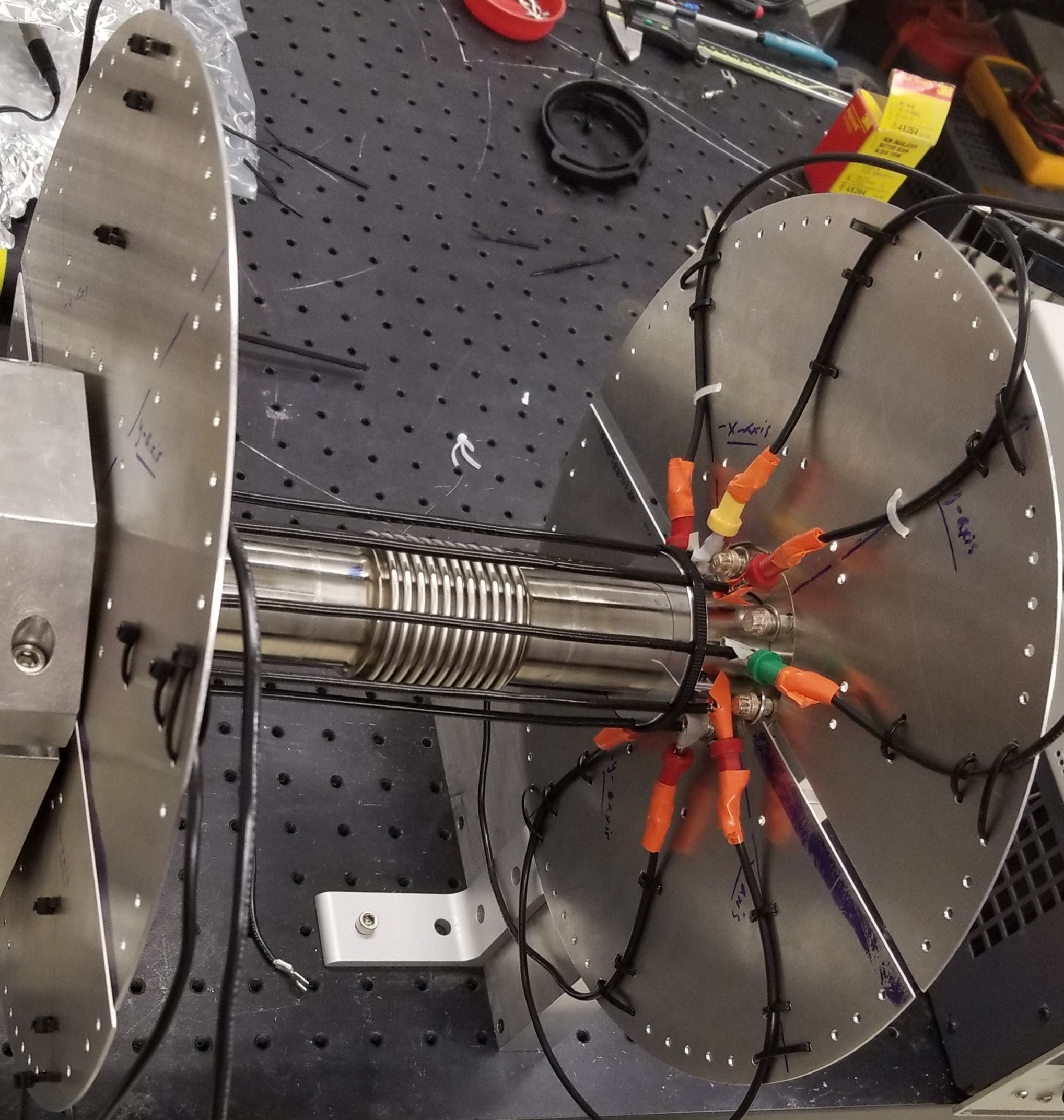}
  \includegraphics[scale=0.035]{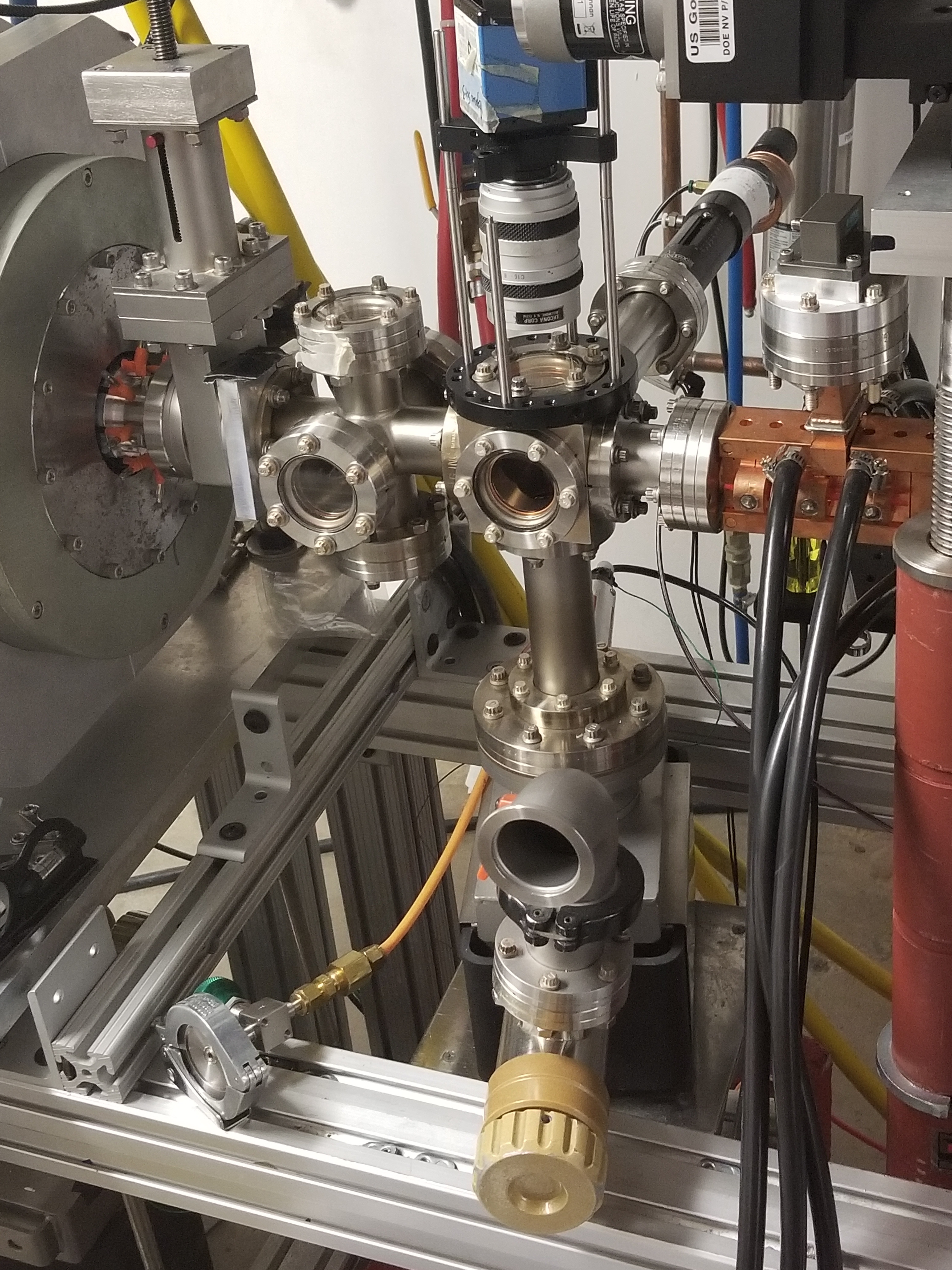}
  \captionsetup{justification=centerlast}
  \caption{(Left) The beam pipe on which the quadrupole configurations are strapped to. (Right) SQC placed inside the emittance compensation solenoid.}
  \label{image_corrector}
\end{figure}

Now we can relate Eqns. \ref{B_currents} and \ref{B_gradient} to identify the currents needed to obtain the target quadrupole:
 \begin{align}
     I_1=\frac{\pi R^2 G}{2\mu_0}\cos (2\alpha) \\
     I_2=\frac{\pi R^2 G}{2\mu_0}\sin (2\alpha)
 \end{align}
The corrector device for the Pegasus beamline had a 30mm radii and 240mm length. It is shown in Fig. \ref{image_corrector}.

\subsection{Emittance Measurements}
\begin{figure*}[t]
    \centering
     \includegraphics[scale=0.45]{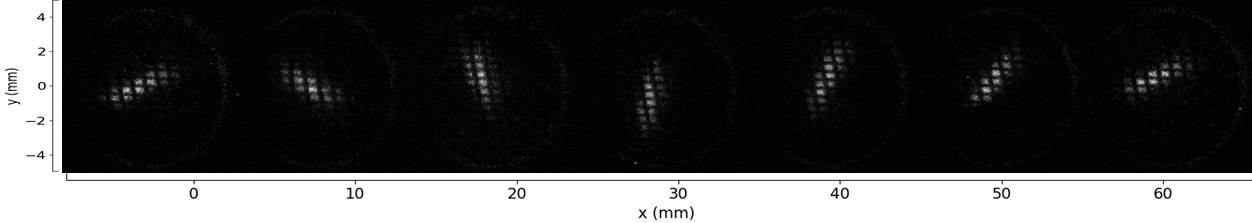}
     \caption{Images from Yag screen taken with ICCD camera for different SQC angles.}
    \end{figure*}
The SQC was placed inside the rotating Larmor frame of the solenoid, so the results from section \ref{emit_corrsolenoid} do apply directly. We resorted to General Particle Tracer (GPT) simulations of the beamline to determine how the final emittance correction shifted as a function of the Larmor angle.

The parameters used in the simulations and in the experiments are reported in Table 1. The error quadrupole location, angle and strength reported in the table are obtained by matching the simulation output to the experimental measurements and should be viewed as effective quantities. 

\begin{table}[b]
\caption{\label{parameters} Parameters for SQC measurement and simulations.}
\begin{ruledtabular}
\begin{tabular}{lcdr}
\textrm{Parameter}&
\textrm{Value}\\
\colrule
Gun gradient & 66.7 MV/m \\
Injection phase &  \ang{19.5}\\
RMS Laser spot size at the cathode& 25$\mu$m \\
RMS Laser pulse length & 100 fs\\
E-beam energy & 3.65 MeV\\
E-beam charge & 100 fC\\
Solenoid field (peak) & 0.18 T \\
Solenoid effective length & 0.15 m \\
Solenoid Larmor Angle & \ang{47} \\
Solenoid location & 0.286m\\
Error quadrupole location downstream & 0.11m\\
Error quadrupole focal length & 0.96m \\
Error quadrupole angle & \ang{30} \\
\end{tabular}
\end{ruledtabular}
\end{table}

\begin{figure}[]
\includegraphics[scale=0.45]{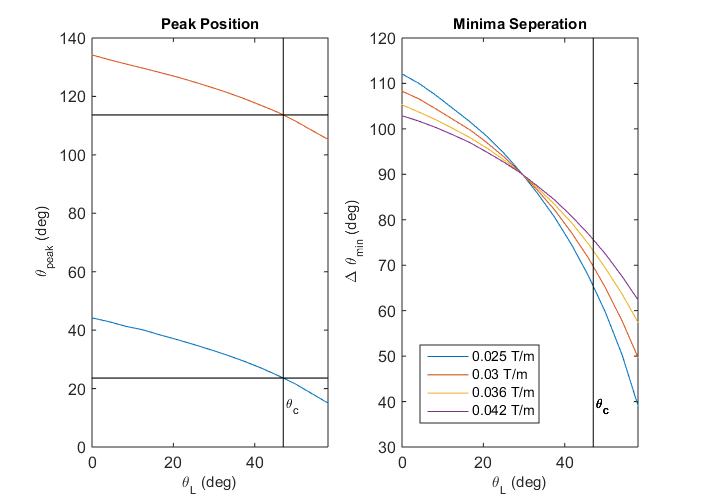}
\captionsetup{justification=centerlast}
  \caption{(Left) Peak positions plotted as a function of Larmor angle.(Right) Emittance minima separation plotted as a function of emittance compensation solenoid Larmor angle.}
  \label{gpt_larmorshift}
\end{figure}

The Larmor rotation angle of the Pegasus emittance compensation solenoid varies between 40 and 50 degrees when used to focus the beam before the quadrupole triplet used in grid based shadowgraph emittance measurements. We show precisely how the Larmor rotation shifts the emittance compensation curves when the SQC is centered inside the solenoid in Fig. \ref{gpt_larmorshift}. The peak and minima positions shift as the solenoid is dialed from 0 up to and beyond operational values. The peaks on the emittance contour for zero magnetic field in the solenoid start at 45 and 135 degrees respectively (consistently with Fig. 2 and 4) and as a function of Larmor angle simply change in phase with each other by approximately half of the rotation angle. The separation between minima shrinks as a function of the Larmor angle and would vanish as the strength of the solenoid is further increased. Beyond that value, the minimum compensation emittance values also increase. Note though that solenoid strength is typically fixed in a given beamline configuration. These plots serve as a guide to interpret the experimental results reported below, where the Larmor angle corresponding to the solenoid magnetic field used was $\theta_L=47^{\circ}$, and the minimum emittances are found near a correction angle of $80^{\circ}$.

Single-shot 4D beam matrix TEM grid shadowgraph measurements were performed to reconstruct the beam matrix at a screen located 3.95m downstream of the cathode. The goal was to scan the angle and gradient of the corrector device to achieve minimization of beam coupling correlations and validate our understanding of the compensation mechanism. The TEM grid shadowgraphy technique has shown suitable for retrieving with high fidelity the second order moments providing sufficient transmission for contrast for low charge beams ($<$~pC). However, the measurement of trace space emittances is particularly sensitive to the screen point spread function \cite{Marx_TEMgrids}, due to the strong correlation in phase space required by the point projection geometry of the measurement. Sample images are presented in Fig. 6, where it can be seen as the SQC angle is adjusted through $180^{\circ}$, the beam envelope also rotates through the same angular range.

The resulting data were compared with GPT simulations and found to be consistent with an effective error quadrupole with a 1~m focal length quadrupole oriented at 30 degrees located at the exit of the RF gun.
\begin{figure}[]
  \includegraphics[scale=0.5]{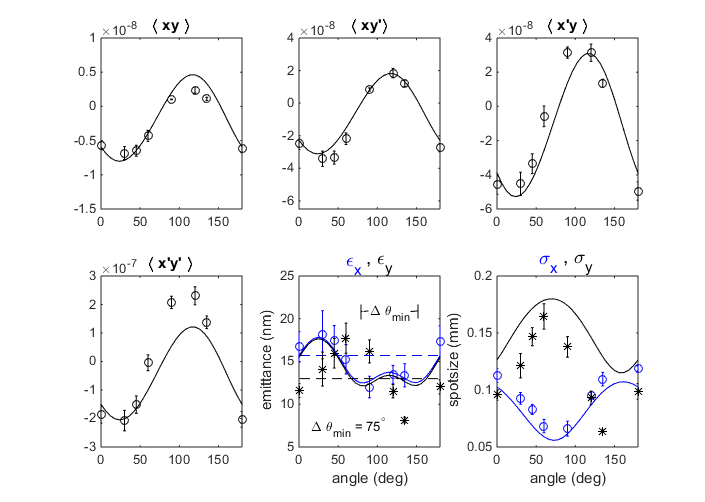}
  \captionsetup{justification=centerlast}
  \caption{Reconstruction of beam matrix and projected emittance compared to simulation of PEGASUS beamline as the SQC angle is dialed. The reconstruction data is overlayed with simulation (solid lines), and the emittance values measured with the SQC off (dashed lines). The error bars represent the standard deviation of the fluctuations over the course of 30 shots.}
  \label{measurements}
\end{figure}
Fig. \ref{measurements} shows both measured and numerically obtained off-diagonal beam matrix elements, projected emittances, and spot sizes as a function of the rotation angle. The dashed lines are the projected emittance values measured with the SQC off ($\epsilon_x=16\pm1$nm rad, $\epsilon_y=13\pm 1$nm rad). There is a small discrepancy in the projected emittances which can be explained if we relax the initial assumption of round beam before the error quadrupole. If we allow for a slight asymmetry in the x and y trace spaces at the cathode, simulations would be in better agreement with the data.

Looking at the beam matrix elements, it is easy to note that the coupling terms are minimized when the projected emittances are at a minimum. For the perfect compensation case, this will obviously be exact, but even in the experimental case where the phase advance is not zero, it is approximately true that the off-diagonal coupling elements are minimized by an optimal choice of corrector settings. This observation provides a very practical guideline to follow in the optimization, which is to tune the angle of the SQC to minimize the $\langle xy \rangle$ moment. In Fig. 6, the central image corresponds with the smallest emittance we measured.
 
\section{Summary and Conclusion}
We have presented an analysis of projected emittance growth from undesirable quadrupole moments in photoinjectors and how to correct them. The complete solution to the zero-phase advance case was used as a starting point in the discussion. It was demonstrated that non-zero phase advance between an error quad and the corrector hinders the ability to completely correct the coupling correlation and retrieve the initial projected emittances. However, a complete correction was found again when the beam distribution at the entrance to the error quad is imaged at the corrector. 

An experiment was carried out at the UCLA PEGASUS beamline where a corrector comprised of two indendent quadrupole wire configurations was fastened to the vacuum pipe inside the solenoid at the gun exit.  Emittance measurements were taken as a function of correction angle. A good agreement between simulations and the analytical prediction was found once the Larmor rotation shift of the solenoid was taken into account. 

We conclude with a possible prescription for tuning an SQC. Quadrupole moments mainly arise in either the RF-gun or emittance compensation solenoid immediately following it. The phase advance between these components is usually small enough that the zero-phase advance approach reviewed in the first section is well suited to guide a beamline operator. The gradient of the corrector should be chosen to be approximately equal to the presumed error quadrupoles, then one can adjust the SQC angle until $\langle xy \rangle=0$ to obtain minimal coupling between the trace spaces.

\bibliography{main}

\end{document}